\documentclass[12pt]{article}
\usepackage{amsmath,amssymb,amsthm,amsxtra,overpic,bbm,bm,epsfig,subfigure}
\usepackage{mathrsfs}
\usepackage{graphicx}
\usepackage{color}
\usepackage{comment}
\usepackage{epstopdf}
\usepackage{float}
\usepackage{cite}
\usepackage{hyperref}
\usepackage{slashed,stmaryrd}

\def\thefootnote{\fnsymbol{footnote}}

\begin{document}

\vspace{0.2cm}

\begin{center}
{\Large\bf Impact of an eV-mass sterile neutrino on the neutrinoless double-beta decays: a Bayesian analysis}
\end{center}

\vspace{0.2cm}

\begin{center}
{\bf Guo-yuan Huang} \footnote{E-mail: huanggy@ihep.ac.cn},
\quad
{\bf Shun Zhou} \footnote{E-mail: zhoush@ihep.ac.cn}
\\
\vspace{0.2cm}
{\small Institute of High Energy Physics, Chinese Academy of
Sciences, Beijing 100049, China \\
School of Physical Sciences, University of Chinese Academy of Sciences, Beijing 100049, China}
\end{center}

\vspace{1.5cm}

\begin{abstract}
To quantitatively assess the impact of an eV-mass sterile neutrino on the neutrinoless double-beta ($0\nu \beta \beta$) decays, we calculate the posterior probability distribution of the relevant effective neutrino mass $|m^\prime_{ee}|$ in the (3+1)$\nu$ mixing scenario, following the Bayesian statistical approach. The latest global-fit analysis of neutrino oscillation data, the cosmological bound on the sum of three active neutrino masses from {\it Planck}, and the constraints from current $0\nu\beta\beta$ decay experiments are taken into account in our calculations. Based on the resultant posterior distributions, we find that the average value of the effective neutrino mass is shifted from $\overline{|m^{}_{ee}|} = 3.37\times 10^{-3}~{\rm eV}$ (or $7.71\times 10^{-3}~{\rm eV}$) in the standard 3$\nu$ mixing scenario to $\overline{|m^{\prime}_{ee}|}=2.54\times 10^{-2}~{\rm eV}$ (or $2.56\times 10^{-2}~{\rm eV}$) in the (3+1)$\nu$ mixing scenario, with the logarithmically uniform prior on the lightest neutrino mass (or on the sum of three active neutrino masses). Therefore, a null signal from the future $0\nu\beta\beta$ decay experiment with a sensitivity to $|m^{}_{ee}| \approx  \mathcal{O}(10^{-2}_{})~{\rm eV}$ will be able to set a very stringent constraint on the sterile neutrino mass and the active-sterile mixing angle.
\end{abstract}


\def\thefootnote{\arabic{footnote}}
\setcounter{footnote}{0}

\newpage

\section{Introduction}
Whether massive neutrinos are Majorana or Dirac particles is one of the most important problems in particle physics~\cite{Majorana:1937vz,Racah:1937qq, Tanabashi:2018oca}. Quite a number of neutrinoless double-beta ($0\nu\beta\beta$) decay experiments are devoted to answering this question~\cite{Furry:1939qr, Rodejohann:2011mu, Bilenky:2012qi, Rodejohann:2012xd, Bilenky:2014uka, Pas:2015eia, DellOro:2016tmg}. If massive neutrinos are Majorana particles and thus lepton number violation exists in nature, then the $0\nu\beta\beta$ decays $A(Z, N) \to A(Z+2, N-2) + 2e^-$ could take place in some even-even nuclei, namely, both the proton number $Z$ and the neutron number $N$ for the nuclear isotope $A(Z, N)$ are even. Assuming the exchange of light Majorana neutrinos to be responsible for the $0\nu\beta\beta$ decays, one can find that the half-life of the relevant nuclear isotope is given by~\cite{Rodejohann:2011mu} 
\begin{eqnarray}
(T^{0\nu}_{1/2})^{-1} = G^{}_{0\nu}|\mathcal{M}^{}_{0\nu}|^2 \frac{|m^{}_{ee}|^2}{m^{2}_{e}}\;,
\label{eq:halflife}
\end{eqnarray}
where $G^{}_{0\nu}$ is the phase-space factor, $\mathcal{M}^{}_{0\nu}$ is the nuclear matrix element (NME), and $m^{}_{e}$ is the electron mass. In Eq.~(\ref{eq:halflife}), the effective neutrino mass $|m^{}_{ee}|$ collects the contributions from light Majorana neutrinos involved in the $0\nu\beta\beta$ decays.

In the standard three neutrino ($3\nu$) mixing scenario, the effective neutrino mass is defined as $|m^{}_{ee}| \equiv |m^{}_{1} U^2_{e1}+m^{}_{2}U^2_{e2}+m^{}_{3}U^2_{e3}|$, where the absolute neutrino masses $m^{}_i$ and the lepton flavor mixing matrix elements $U^{}_{ei}$ (for $i = 1, 2, 3$) appear. When the conventional parametrization of the flavor mixing matrix $U$ is adopted~\cite{Tanabashi:2018oca}, i.e., $U^{}_{e1} = \cos \theta^{}_{13} \cos \theta^{}_{12} e^{{\rm i}\rho/2}$, $U^{}_{e2} = \cos \theta^{}_{13} \sin \theta^{}_{12}$ and $U^{}_{e3} = \sin \theta^{}_{13} e^{{\rm i}\sigma/2}$, we have
\begin{eqnarray}
m^{}_{ee} \equiv m^{}_{1} \cos^2\theta^{}_{13} \cos^2\theta^{}_{12} e^{{\rm i} \rho} + m^{}_{2} \cos^2\theta^{}_{13} \sin^2\theta^{}_{12} + m^{}_{3} \sin^2\theta^{}_{13} e^{{\rm i} \sigma} \;,
\label{eq:mee}
\end{eqnarray}
where $\{\theta^{}_{12}, \theta^{}_{13}\}$ are two of three neutrino mixing angles, and $\{\rho, \sigma\}$ are the Majorana-type CP-violating phases. Note that $m^{}_2$ is nonzero no matter whether the normal neutrino mass ordering (NO) with $m^{}_1 < m^{}_2 < m^{}_3$ or the inverted neutrino mass ordering (IO) with $m^{}_3 < m^{}_1 < m^{}_2$ is considered. Therefore, such a parametrization is favorable in the discussions about the limiting case of $m^{}_1 \to 0$ (for NO) or $m^{}_3 \to 0$ (for IO), for which one of two Majorana-type CP violating phases just disappears together with the lightest neutrino mass. 

However, if the eV-mass sterile neutrino indeed exists as a solution to the anomalies in the short-baseline neutrino experiments~\cite{Giunti:2019aiy, Aguilar:2001ty, AguilarArevalo:2008rc, Aguilar-Arevalo:2018gpe, Giunti:2010zu, Mention:2011rk, Abdurashitov:2009tn, Kaether:2010ag}, it will contribute as well to the $0\nu \beta \beta$ decays. In this case, the effective neutrino mass is given by $|m^{\prime}_{ee}| \equiv |m^{}_{1} V^2_{e1} + m^{}_{2} V^2_{e2} + m^{}_{3} V^2_{e3} + m^{}_{4} V^2_{e4}|$ with $m^{}_{4}$ being the mass of the sterile neutrino and $V^{}_{ei}$ (for $i=1,2,3,4$) being the first-row elements of the mixing matrix in the (3+1)$\nu$ mixing scenario. Adopting the standard parametrization of the mixing matrix, one can express the effective neutrino mass as
\begin{eqnarray}
|m^{\prime}_{ee}| & \equiv &  | m^{}_{ee} \cos^2\theta^{}_{14} + m^{}_{4}\sin^2\theta^{}_{14} e^{i \omega}|\;,
\label{eq:meeprime}
\end{eqnarray}
where $m^{}_{ee}$ takes the same form as in Eq.~(\ref{eq:mee}), $\theta^{}_{14}$ is the active-sterile neutrino mixing angle, and $\omega$ is the additional Majorana-type CP-violating phase. Using the best-fit values $\Delta m^2_{41} \equiv m^2_4 - m^2_1 = 1.7~{\rm eV}^2$ and $\sin^2 \theta^{}_{14} = 0.019$ from the global-fit analysis of the short-baseline neutrino oscillation data~\cite{Gariazzo:2017fdh, Dentler:2018sju}, one can find that the contribution from the sterile neutrino $|m^{}_4 \sin^2 \theta^{}_{14}| \approx 2.5\times 10^{-2}~{\rm eV}$ can be comparable to that from active neutrinos $|m^{}_{ee}| \lesssim 0.1~{\rm eV}$, which is constrained by the cosmological observations~\cite{Aghanim:2018eyx} and current $0\nu\beta\beta$ decay experiments~\cite{Albert:2014awa, KamLAND-Zen:2016pfg, Agostini:2017iyd, Alduino:2017ehq, Aalseth:2017btx, Agostini:2018tnm}. 

With a ton-scale target mass, the future $0\nu\beta\beta$ experiments will be able to probe $|m^{}_{ee}|$ to the $\mathcal{O}(10^{-2})~{\rm eV}$ level \cite{Agostini:2017jim}, covering the whole range of $|m^{}_{ee}|$ in the IO case. However, in the NO case, the effective neutrino mass can be as small as $|m^{}_{ee}| \approx (1.6 \cdots 3.6) \times 10^{-3} ~{\rm eV}$ when the lightest neutrino mass $m^{}_{1}$ is vanishing, or even vanishing in the contrived region of parameter space when the cancellation among the contributions from different neutrino mass eigenstates occurs~\cite{Xing:2003jf, Xing:2015zha, Xing:2016ymd}. Moreover, the latest global-fit analysis of neutrino oscillation data~\cite{deSalas:2017kay,Capozzi:2018ubv,Esteban:2018azc} does show a preference for the NO at the $3\sigma$ confidence level (C.L.), it is worrisome that $|m^{}_{ee}|$ may be out of the reach of the next generation $0\nu \beta\beta$ decay experiments. To quantitatively assess how likely $|m^{}_{ee}|$ is small, the authors of Refs.~\cite{Agostini:2017jim, Caldwell:2017mqu} have carried out a Bayesian analysis and obtained the posterior distribution of $|m^{}_{ee}|$, given the neutrino oscillation data, current experimental upper bounds on $|m^{}_{ee}|$ and the cosmological bound on the sum of three neutrino masses. For the earlier relevant works, see Refs.~\cite{Benato:2015via, Zhang:2015kaa, Ge:2016tfx}. Although the impact of an eV-mass sterile neutrino on the effective neutrino mass $|m^\prime_{ee}|$ has been considered in Refs.~ \cite{Goswami:2005ng, Goswami:2007kv, Barry:2011wb, Li:2011ss, Girardi:2013zra, Guzowski:2015saa, Giunti:2015kza, Ge:2017erv, Liu:2017ago}, a statistical assessment is still lacking. Therefore, we are motivated to perform a Bayesian analysis of $|m^\prime_{ee}|$ in this work by using the global-fit results of neutrino oscillation data and other available information on the absolute neutrino masses.

The rest of the present paper is organized as follows. In Section 2, we describe the necessary information for the Bayesian analysis. The prior information can be extracted from the global-fit analysis of neutrino oscillation data~\cite{Capozzi:2018ubv, Gariazzo:2017fdh}, the cosmological observations~\cite{Aghanim:2018eyx} and the existing $0\nu\beta\beta$ decay experiments~\cite{Albert:2014awa, KamLAND-Zen:2016pfg, Agostini:2017iyd, Alduino:2017ehq}. Then, the posterior distribution of the standard effective neutrino mass $|m^{}_{ee}|$ and that of $|m^\prime_{ee}|$ are presented in Section 3. Two-dimensional posterior probability densities in the $|m^{\prime}_{ee}|$-$m^{}_{\rm L}$ plane and those in the $|m^{\prime}_{ee}|$-$\rho$ plane have also been given, where $m^{}_{\rm L}$ denotes the lightest neutrino mass. Finally, we make some concluding remarks in Section 4.

\section{The Bayesian Analysis}

The Bayesian analysis provides us with a reasonable statistical framework to update the probability distribution of model parameters in light of the new experimental data. The posterior distribution of model parameters can be obtained according to the Bayesian theorem~\cite{Skilling:book} 
\begin{eqnarray} \label{eq:Bayesian}
P(\Theta,\mathcal{H}^{}_{i}|\mathcal{D}) = \frac{\mathcal{L}(\mathcal{D}|\Theta, \mathcal{H}^{}_{i})\mathcal{\pi}(\Theta,\mathcal{H}^{}_{i})}{\sum^{}_{i}\mathcal{Z}^{}_{i}}\;,
\end{eqnarray}
where $\Theta$ denotes the set of model parameters, $\mathcal{D}$ stands for the available experimental data, and $\{\mathcal{H}^{}_{i}\}$ are the hypotheses or models with $i$ being the model index. Here $\mathcal{L}(\mathcal{D}|\Theta, \mathcal{H}^{}_{i})$ is the likelihood of the data $\mathcal{D}$, assuming the model $\mathcal{H}^{}_{i}$ with the parameters $\Theta$, $\mathcal{\pi}(\Theta,\mathcal{H}^{}_{i})$ is the prior distribution of $\Theta$, and $\mathcal{Z}^{}_{i}$ is the evidence. The evidence $\mathcal{Z}^{}_{i}$ is given by 
\begin{eqnarray} \label{eq:ZEvidence}
\mathcal{Z}^{}_{i} = \int \mathcal{L}(\mathcal{D}|\Theta, \mathcal{H}^{}_{i})\mathcal{\pi}(\Theta,\mathcal{H}^{}_{i}) d^{N}\Theta\;,
\end{eqnarray}
which measures the compatibility of the model with the data, and $N$ is just the dimension of the parameter space. The hypotheses relevant for our analysis are $\mathcal{H}^{}_{\rm NO}$ for the NO and $\mathcal{H}^{}_{\rm IO}$ for the IO in the $3\nu$ or (3+1)$\nu$ mixing scenario. 
The model parameters in the (3+1)$\nu$ mixing scenario include: (i) the involved neutrino oscillation parameters $\{ \sin^2\theta^{}_{13}, \sin^2\theta^{}_{12}, \sin^2\theta^{}_{14},\Delta m^{2}_{\rm sol},\Delta m^{2}_{\rm atm},\Delta m^{2}_{41} \}$, where $\Delta m^2_{\rm sol} \equiv m^2_2 - m^2_1$ and $\Delta m^2_{\rm atm} \equiv m^2_3 - (m^2_2 + m^2_1)/2$ are two mass-squared differences of ordinary neutrinos; (ii) the lightest neutrino mass $m^{}_{\rm L}$, which is $m^{}_{1}$ for $\mathcal{H}^{}_{\rm NO}$ and $m^{}_{3}$ for $\mathcal{H}^{}_{\rm IO}$; (iii) the Majorana-type CP-violating phases $\{\rho,\sigma,\omega\}$; (iv) the phase-space factor and the nuclear matrix element $\{ G^{}_{0\nu}, |\mathcal{M}^{}_{0\nu}| \}$ for the $0\nu\beta\beta$ decays. The overall likelihood function can be constructed as $\mathcal{L} = \mathcal{L}^{}_{\rm 3\nu} \times \mathcal{L}^{}_{\rm cosmo} \times \mathcal{L}^{}_{\rm 0\nu\beta\beta} \times \mathcal{L}^{}_{\rm sterile}$, and the details of the individual likelihood function are summarized as follows.
\begin{itemize}
\item $\mathcal{L}^{}_{\rm 3\nu}$: the likelihood function of the $3\nu$ mixing parameters $\{ \sin^2\theta^{}_{13},  \sin^2\theta^{}_{12}, \Delta m^{2}_{\rm sol}, \Delta m^{2}_{\rm atm} \}$. Given the $\Delta \chi^2$ function from the global-fit analysis in Ref.~\cite{Capozzi:2018ubv}, we can fix the likelihood function $\mathcal{L}^{}_{\rm 3\nu} = \exp(-\Delta \chi^2/2)$, where $\Delta \chi^2$ is defined as 
\begin{eqnarray}
\Delta \chi^2 \equiv \sum^{}_{i} \frac{(\Theta^{}_{i}-\Theta^{\rm bf}_{i})^2}{\sigma^{2}_{i}}\;,
\end{eqnarray}
with $\Theta^{}_{i}$ running over $\{ \sin^2\theta^{}_{13}, \sin^2 \theta^{}_{12}, \Delta m^{2}_{\rm sol}, \Delta m^{2}_{\rm atm} \}$, $\Theta^{\rm bf}_{i}$ the corresponding best-fit value from the global analysis, and $\sigma^{}_{i}$ the symmetrized $1\sigma$ error. See Table.~1 of Ref.~\cite{Capozzi:2018ubv} for more details about the global-fit results of neutrino oscillation data. To be explicit, we list the best-fit values and the corresponding symmetrized $1\sigma$ errors as below
\begin{eqnarray}
&& \sin^2\theta^{}_{12} = (3.04\pm 0.14) \times 10^{-1} \; , \quad  \Delta m^2_{\rm sol} = (7.34 \pm 0.16) \times 10^{-5} ~{\rm eV}^2  \; ,
\nonumber \\
&& \sin^2\theta^{}_{13} = (2.14 \pm 0.08) \times 10^{-2}\; , ~~\quad
\Delta m^2_{\rm atm} = (2.455 \pm 0.034) \times 10^{-3} ~{\rm eV}^2 \; ,\hspace{0.6cm}
\end{eqnarray}
for $\mathcal{H}^{}_{\rm NO}$; and 
\begin{eqnarray}
&& \sin^2\theta^{}_{12} = (3.03\pm 0.14) \times 10^{-1} \; , \quad  \Delta m^2_{\rm sol} = (7.34 \pm 0.16) \times 10^{-5} ~{\rm eV}^2 \; ,
\nonumber \\
&& \sin^2\theta^{}_{13} = (2.18 \pm 0.08)\times 10^{-2} \; , \quad \Delta m^2_{\rm atm} = (-2.441 \pm 0.034) \times 10^{-3} ~{\rm eV}^2 \; ,\hspace{0.6cm}
\end{eqnarray}
for $\mathcal{H}^{}_{\rm IO}$. The latest neutrino oscillation data favor the NO over the IO at the $3\sigma$ level, i.e., the difference between the minima of $\chi^2$ in these two cases is $\Delta \chi^2_{\rm min} \equiv \chi^{\rm IO}_{\rm min} - \chi^{\rm NO}_{\rm min} \approx 9$. The preference for the NO arises mainly from two different data sets. First, the excess of $\nu^{}_e$-like events in the multi-GeV energy range in Super-Kamiokande atmospheric neutrino data can be accommodated by the resonant enhancement of the oscillation probability in the $\nu^{}_\mu \to \nu^{}_e$ channel, leading to $\Delta \chi^2_{\rm min} \approx 4$. Second, the running long-baseline accelerator experiments T2K and NO$\nu$A prefer the value of $\theta^{}_{13}$ that is slightly larger than the precisely measured value from reactor neutrino experiments. Such a tension between accelerator and reactor neutrino experiments will be relieved in the NO case, contributing another $\Delta \chi^2_{\rm min} \approx 4$ to the mass ordering discrimination. To be conservative, we will take $\Delta\chi^2_{\rm min} = 4$ as the preference for the NO over the IO from neutrino oscillation data.
\begin{figure}[t!]
\begin{center}
\hspace{-0.2cm}
\includegraphics[width=0.48\textwidth]{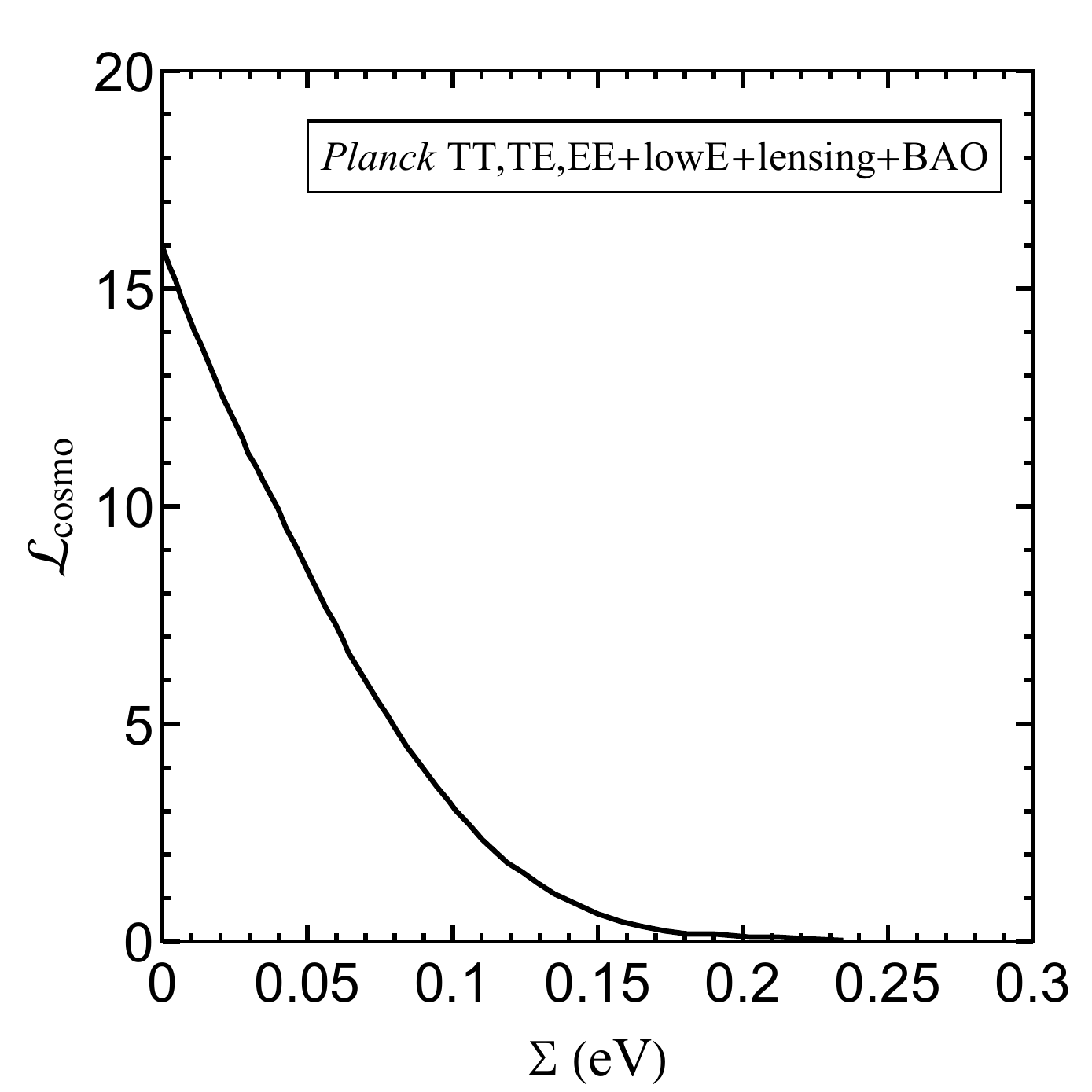} 
\end{center}
\vspace{-0.5cm}
\caption{The likelihood function $\mathcal{L}^{}_{\rm cosmo}$ for the sum of three neutrino masses $\Sigma \equiv m^{}_1 + m^{}_2 + m^{}_3$ from cosmological observations, which has been derived by combining the $Planck ~{\rm TT}, {\rm TE}, {\rm EE} + {\rm lowE} + {\rm lensing} + {\rm BAO}$ data sets~\cite{Aghanim:2018eyx}.}
\label{fig:1}
\end{figure}

\item $\mathcal{L}^{}_{\rm cosmo}$: the likelihood function for the cosmological observations on the sum of three neutrino masses $\Sigma \equiv m^{}_{1}+m^{}_{2}+m^{}_{3}$. After combining several different sets of cosmological data ($Planck ~{\rm TT}, {\rm TE}, {\rm EE} + {\rm lowE} + {\rm lensing} + {\rm BAO}$), the {\it Planck} Collaboration has recently updated the upper limit on the sum of neutrino masses as $\Sigma < 0.12~{\rm eV}$ at the $95\%$ C.L.~\cite{Aghanim:2018eyx}. We obtain the likelihood information by making use of the Markov chain file available from the Planck Legacy Archive (PLA)~\footnote{This is based on the observations with {\it Planck} (\url{http://www.esa.int/Planck}), an ESA science mission with instruments and contributions directly funded by ESA Member States, NASA, and Canada.}. The likelihood function of $\Sigma$ is produced and shown in Fig.~\ref{fig:1} by marginalizing over the other cosmological parameters. Although the sampling file given by PLA has assumed a degenerate mass spectrum of neutrinos, a more solid analysis with the realistic neutrino mass spectrum should not change the result much~\cite{Hannestad:2016fog}. For this reason, the likelihood shown in Fig.~ \ref{fig:1} will be used in the following discussions.

\item $\mathcal{L}^{}_{\rm 0\nu\beta\beta}$: the likelihood function derived from the experimental constraints on the effective neutrino mass $|m^{}_{ee}|$ or $|m^{\prime}_{ee}|$ due to the existing searches for $0\nu\beta\beta$ decays. For simplicity, we implement the likelihood function available from Refs.~\cite{Caldwell:2017mqu,Alduino:2017ehq} in our analysis. Although both $\mathcal{L}^{}_{\rm 0\nu\beta\beta}$ and $\mathcal{L}^{}_{\rm cosmo}$ contain the information about the absolute scale of neutrino masses, the constraint on $|m^{}_{ee}|$ from the $0\nu\beta\beta$ decays suffers from a large theoretical uncertainty in the prediction for the NME. For instance, the tightest bound comes from the KamLAND-Zen experiment~\cite{KamLAND-Zen:2016pfg}, namely, $|m^{}_{ee}| \lesssim (61\cdots 165)~{\rm meV}$. Given further uncertainties from the mixing parameters and the unknown Majorana CP-violating phases, the $0\nu\beta\beta$ decays are not so informative about the absolute scale of neutrino masses when compared to the cosmological observations. 

\item $\mathcal{L}^{}_{\rm sterile}$: the likelihood function encoding the global-fit analysis of sterile neutrino mass and mixing parameters $\{ \theta^{}_{14}, \Delta m^{2}_{41}\}$. In practice, we determine the likelihood function as $\mathcal{L}^{}_{\rm sterile} = \exp[ -\Delta \chi^2_{\rm sterile}(\theta^{}_{14}, \Delta m^{2}_{41}) /2]$ by using the $\Delta \chi^2$ distribution in Fig.~9 of Ref.~\cite{Gariazzo:2017fdh}. The result of the so-called pragmatic 3+1 global fit “PrGlo17” will be utilized~\cite{Gariazzo:2017fdh}, where the tension between appearance and disappearance oscillation data can be somewhat relaxed by ignoring the excess of low-energy ${\nu}^{}_{e}$-like events observed in the MiniBooNE experiment.
\end{itemize}

\begin{figure}[t!]
\begin{center}
\subfigure{
\hspace{-0.2cm}
\includegraphics[width=0.48\textwidth]{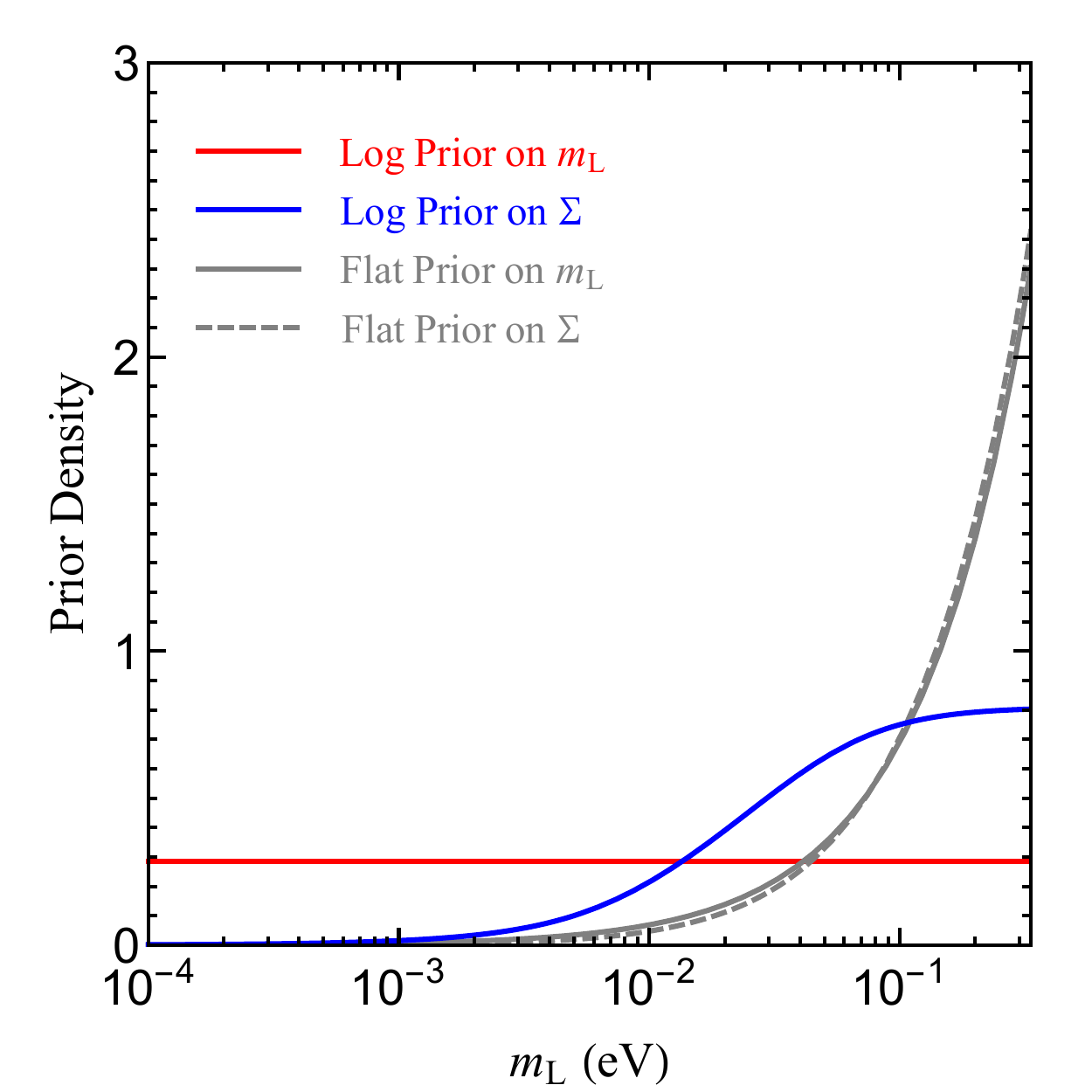} }
\subfigure{
\includegraphics[width=0.48\textwidth]{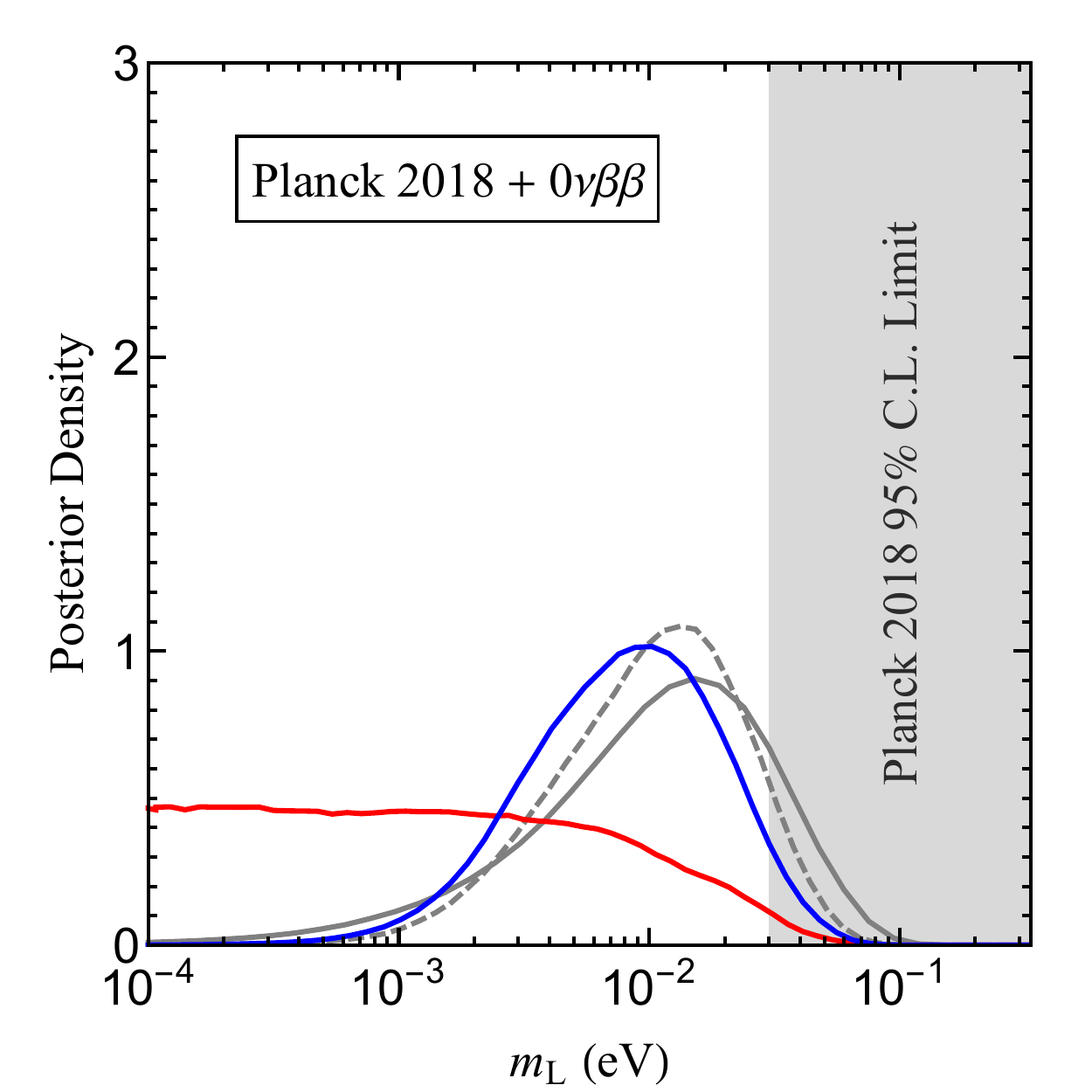} }
\end{center}
\vspace{-0.5cm}
\caption{The distribution functions of the prior probability (\emph{Left Panel}) and the posterior probability (\emph{Right Panel}) for the lightest neutrino mass $m^{}_{\rm L}$. The prior choices include the logarithmic prior on $m^{}_{\rm L}$ (red curve), the logarithmic prior on $\Sigma$ (blue curve), the flat prior on $m^{}_{\rm L}$ (solid gray curve), and the flat prior on $\Sigma$ (dashed gray curve). Their posterior distributions after considering the {\it Planck}~2018 and the $0\nu\beta\beta$ decay data are given correspondingly in the right panel.}
\label{fig:2}
\end{figure}

After having the likelihood functions constructed from various experimental observations, we need to make clear the prior probability distributions of the model parameters, which reflect our knowledge about them prior to any experimental data. First, neutrino mass-squared differences and mixing angles $\{ \sin^2\theta^{}_{13}, \sin^2\theta^{}_{12}, \sin^2\theta^{}_{14}, \Delta m^{2}_{\rm sol}, \Delta m^{2}_{\rm atm}, \Delta m^{2}_{41} \}$ are assumed to be uniformly distributed in their allowed ranges that are wide enough to cover their global fit results. Since the oscillation data are rather informative, different choices of prior distributions of these parameters do not have much impact on the final posterior distributions. Second, the Majorana CP-violating phases are completely unknown, so it is reasonable to adopt the flat priors in the range of $[0\cdots 2\pi)$. In addition, we have to mention that the prior distributions for the following relevant parameters are by no means unique but will be incorporated into our calculations for practical purposes.
\begin{itemize}
\item As indicated in Eq.~(\ref{eq:halflife}), the phase-space factor $G^{}_{0\nu}$ and the NME $|\mathcal{M}^{}_{0\nu}|$ are needed when we try to translate the experimental constraint on the half-life into that on the effective neutrino mass. The phase-space factors for different nuclear isotopes have been computed in Refs.~\cite{Rodejohann:2011mu, Suhonen:1998ck, Kotila:2012zza}, and we use the central values from Ref.~\cite{Kotila:2012zza}, e.g., $G^{}_{0\nu}({}^{76}{\rm Ge}) = 6.15 \times 10^{-15}~{\rm yr}^{-1}$, $G^{}_{0\nu}({}^{130}{\rm Te}) = 3.70 \times 10^{-14}~{\rm yr}^{-1}$ and $G^{}_{0\nu}({}^{136}{\rm Xe}) = 3.79 \times 10^{-14}~{\rm yr}^{-1}$, which have been obtained with the axial vector coupling constant $g^{}_{\rm A} = 1.27$. We assume that $G^{}_{0\nu}$ can be described by the Gaussian distribution with the aforementioned central value and a relative error of $7\%$. On the other hand, the NME for a specific nuclear isotope encoding the information about the nuclear structure has been theoretically calculated in a variety of nuclear models. The differences among these calculations can be treated as the theoretical uncertainty. We define this uncertainty as $\sigma^{}_{\rm NME} \equiv \sum^{}_{i}(|\mathcal{M}^{i}_{0\nu}|-\overline{|\mathcal{M}^{}_{0\nu}|})^2/n^{}_{\rm NME}$, where $|\mathcal{M}^{i}_{0\nu}|$ is the NME value of the $i$th model, $\overline{|\mathcal{M}^{}_{0\nu}|}$ is the averaged NME value of all models, and $n^{}_{\rm NME}$ is the total number of models. Using the tabulated NME values in Ref.~\cite{Guzowski:2015saa}, we find that $\overline{|\mathcal{M}^{}_{0\nu}|}({}^{76}{\rm Ge},{}^{130}{\rm Te},{}^{136}{\rm Xe}) = (4.88,3.94,2.73) $ 
and $\sigma^{}_{\rm NME}({}^{76}{\rm Ge},{}^{130}{\rm Te},{}^{136}{\rm Xe}) = (1.14,0.90,0.80)$. Then the Gaussian distribution with the central value $\overline{|\mathcal{M}^{}_{0\nu}|}$ and the standard deviation $\sigma^{}_{\rm NME}$ is assumed for each nuclear isotope.

\item For the prior of the lightest neutrino mass $m^{}_{\rm L}$, a more careful study should be performed. Four kinds of prior distributions for $m^{}_{\rm L}$ are usually considered: (i) a logarithmic prior on $m^{}_{\rm L}$ with an adjustable lower cutoff that we choose to be $10^{-4}~{\rm eV}$; (ii) a logarithmic prior on $\Sigma$ with a natural lower cutoff at $0.06~{\rm eV}$ for NO or at $0.1~{\rm eV}$ for IO, as required by neutrino oscillation experiments; (iii) a flat prior on $m^{}_{\rm L}$; (iv) a flat prior on $\Sigma$. The prior probability distributions have been plotted with respect to $\log_{10}(m^{}_{\rm L}/{\rm eV})$ in the left panel of Fig.~\ref{fig:2}, where one can see that the flat priors on $m^{}_{\rm L}$ (gray solid curve) and $\Sigma$ (gray dashed curve) lead to nearly the same distribution. After incorporating the experimental limits from {\it Planck}~2018 and the $0\nu\beta\beta$ decays, as shown in the right panel of Fig.~\ref{fig:2}, we observe that the logarithmic prior on $m^{}_{\rm L}$ (red solid curve) gives rise to a posterior distribution that is very different from those in the other scenarios. This is because a large weight has been given to very small neutrino masses in the former case. In the following discussions, we focus only on two different prior distributions, i.e., the logarithmic prior on $m^{}_{\rm L}$ and the logarithmic prior on $\Sigma$, both of which are scale invariant. Since the posterior distribution of $m^{}_{\rm L}$ with logarithmic prior on $\Sigma$ is very similar to those with two flat priors, the posterior distribution of the effective neutrino mass in the former case should also be roughly applicable to those in the latter two cases. 
\end{itemize}

Finally, we make some comments on the current experimental hint on neutrino mass ordering by combining the data sets of neutrino oscillation experiments, cosmological observations and the $0\nu\beta\beta$ decays, for which the likelihood functions are given by $\mathcal{L}^{}_{\rm 3\nu}$, $\mathcal{L}^{}_{\rm cosmo}$ and $\mathcal{L}^{}_{\rm 0\nu\beta\beta}$, respectively. The preference odds for NO over IO can be represented by the Bayes factor, i.e., $\mathcal{B} \equiv \mathcal{Z}^{}_{\rm NO}/\mathcal{Z}^{}_{\rm IO}$. With the help of Eq.~(\ref{eq:ZEvidence}), one can calculate the evidences for NO and IO and thus their ratio. The dependence of $\mathcal{B}$ on the choice of the $m^{}_{\rm L}$ prior distribution is found to be very weak. Given identical prior information on both mass orderings, we consider only the cosmological observations $\mathcal{L}^{}_{\rm cosmo}$ and obtain the logarithm of the Bayes factor as $\log(\mathcal{B}^{}_{\rm cosmo}) \approx 0.85$~\footnote{Note that the subscript of the Bayes factor ${\cal B}^{}_{\rm cosmo}$ herein refers to the cosmological data that have been used in the calculations, and likewise for the Bayes factors from other data sets and their combinations.}, corresponding to ${\cal B}^{}_{\rm cosmo} \approx 2.34$, which is in concordance with the results from Refs.~\cite{Hannestad:2016fog, Gerbino:2016ehw, Vagnozzi:2017ovm, Capozzi:2017ipn}. If only the $0\nu\beta\beta$ decay experiments are considered, then we get $\log(\mathcal{B}^{}_{ 0\nu\beta\beta}) \approx 0.2$. A combination of the cosmological observations and $0\nu\beta\beta$ decay data leads to $\log(\mathcal{B}^{}_{ {\rm cosmo} + 0\nu\beta\beta}) \approx 1.1$. Regarding the three-flavor neutrino oscillation data, if we take the conservative choice of $\Delta \chi^2_{\rm min} \approx 4$ for two neutrino mass orderings, which has been used to construct ${\cal L}^{}_{3\nu}$, the logarithm of the Bayes factor turns out to be $\log(\mathcal{B}^{}_{ 3\nu})= 2$. Combining ${\cal L}^{}_{\rm cosmo}$, ${\cal L}^{}_{0\nu\beta\beta}$ and ${\cal L}^{}_{3\nu}$ together, one can find the total Bayes factor $\mathcal{B}^{}_{\rm tot} \approx 22$. As we have mentioned before, the global-fit analysis of all the neutrino oscillation data gives rise to a $3\sigma$ preference for the NO, corresponding to ${\cal B}^{}_{3\nu} \approx 90$. If such a stronger preference for the NO is implemented instead of the conservative one, the total Bayes factor from all the data sets becomes $\mathcal{B}^{}_{\rm tot} \approx 270$, showing a strong evidence for the NO according to the Jeffreys scale~\cite{Trotta:2008qt}. The addition of $\mathcal{L}^{}_{\rm sterile}$ into the analysis does not alter the above conclusions, since the short-baseline neutrino oscillation experiments are insensitive to the mass ordering of three ordinary neutrinos.

\section{Posterior Distributions}
\begin{figure}[t!]
\begin{center}
\subfigure{
\hspace{-0.2cm}
\includegraphics[width=0.48\textwidth]{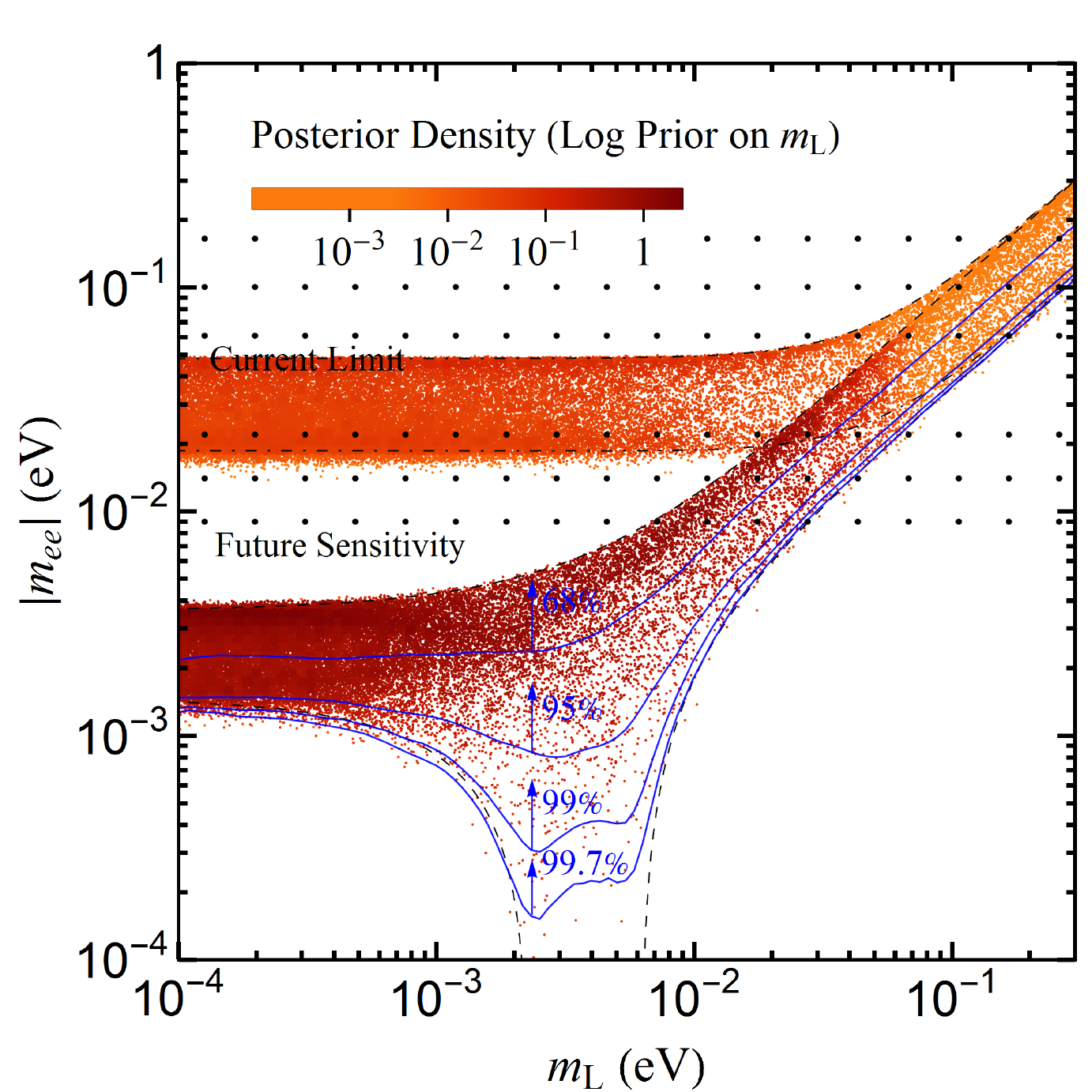} }
\subfigure{
\includegraphics[width=0.48\textwidth]{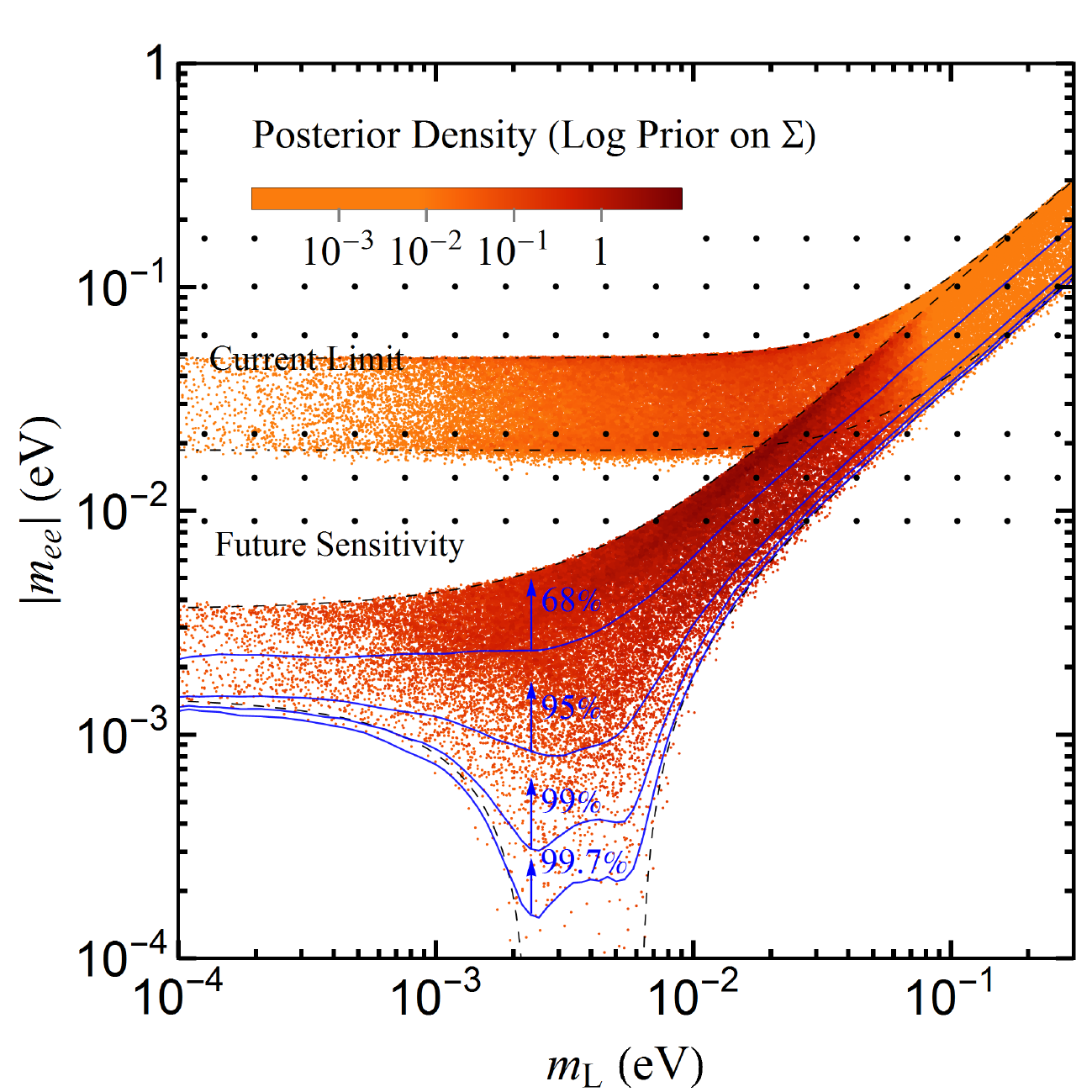} }
\subfigure{
\hspace{-0.2cm}
\includegraphics[width=0.48\textwidth]{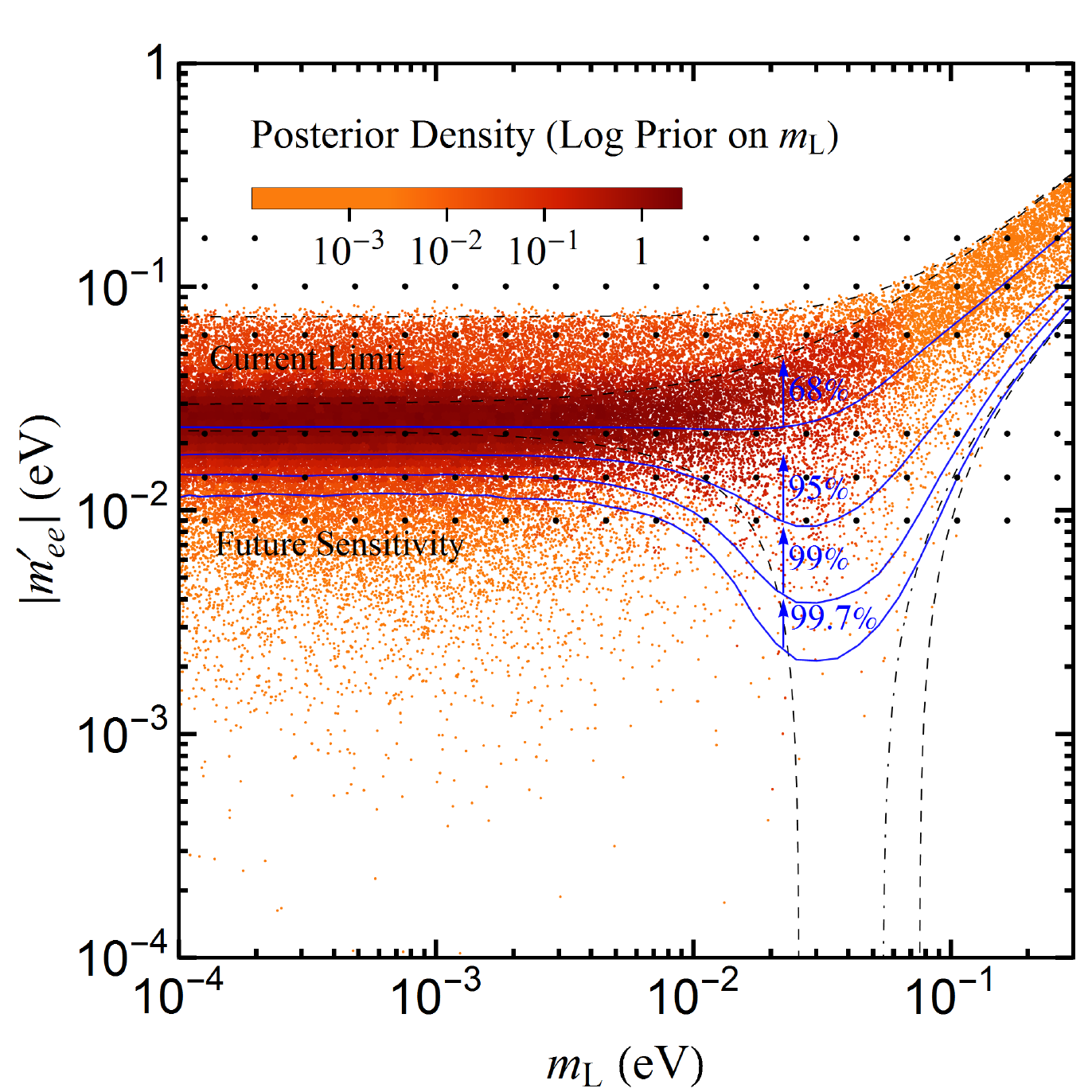} }
\subfigure{
\includegraphics[width=0.48\textwidth]{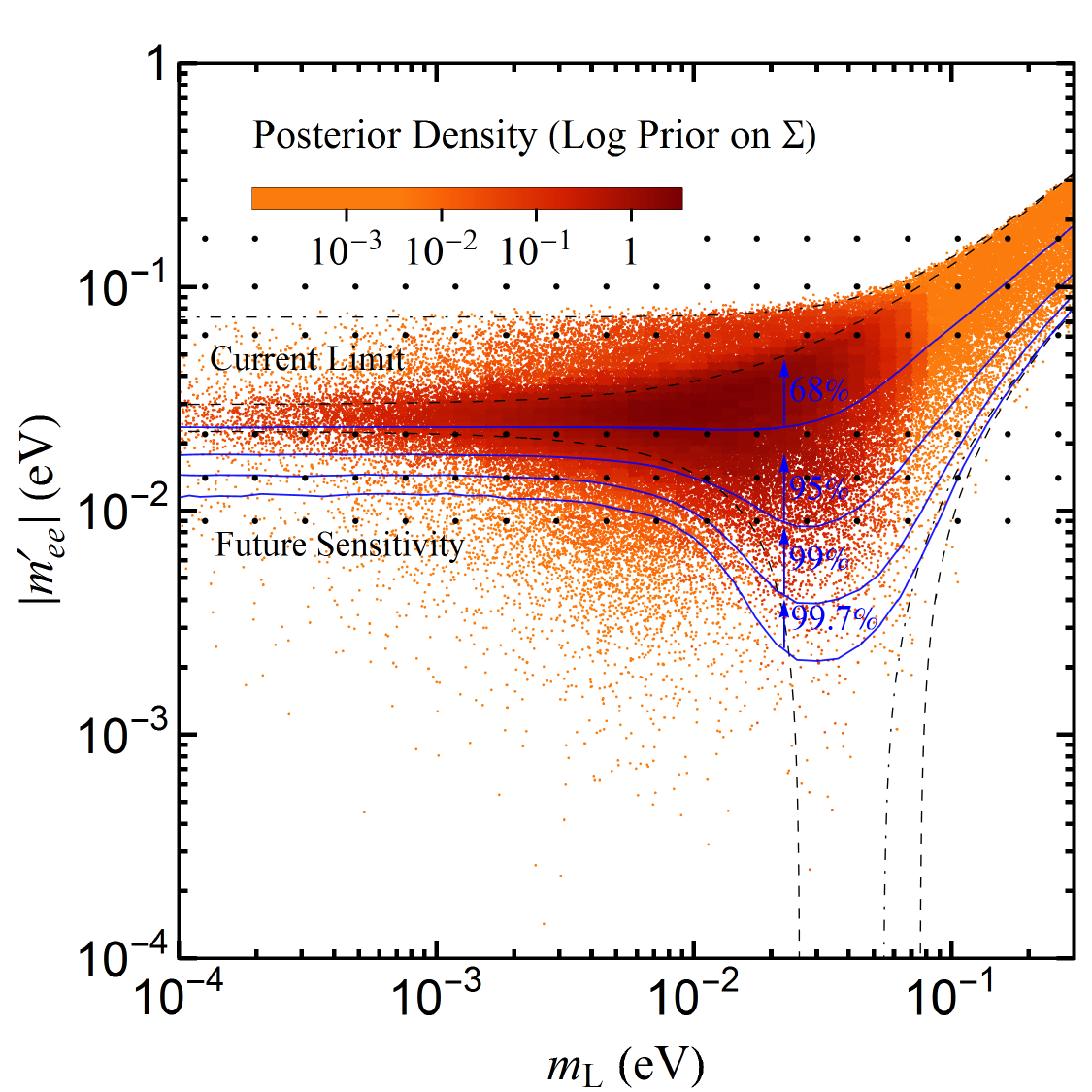} }
\end{center}
\vspace{-0.5cm}
\caption{The posterior probability densities in the $|m^{}_{ee}|$-$m^{}_{\rm L}$ (the upper row) or $|m^{\prime}_{ee}|$-$m^{}_{\rm L}$ (the lower row) plane for different choices of models and priors: (i) the standard 3$\nu$ mixing scenario with the logarithmic prior on $m^{}_{\rm L}$ (\emph{Upper-left Panel}), (ii) the standard 3$\nu$ mixing scenario with the logarithmic prior on $\Sigma$ (\emph{Upper-right Panel}), (iii) the (3+1)$\nu$ mixing scenario with the flat prior on $m^{}_{\rm L}$ (\emph{Lower-left Panel}) and (iv) the (3+1)$\nu$ mixing scenario with the flat prior on $\Sigma$ (\emph{Lower-right Panel}). The blue solid curves stand for the contours of the probability for the true value of the effective neutrino mass to be above a certain $|m^{}_{ee}|$. The dashed and dot-dashed curves represent the boundaries of the effective masses obtained by using the best-fit oscillation parameters and free Majorana CP phases in the NO and IO cases, respectively.}
\label{fig:3}
\end{figure}
\begin{figure}[t!]
\begin{center}
\subfigure{
\hspace{-0.2cm}
\includegraphics[width=0.48\textwidth]{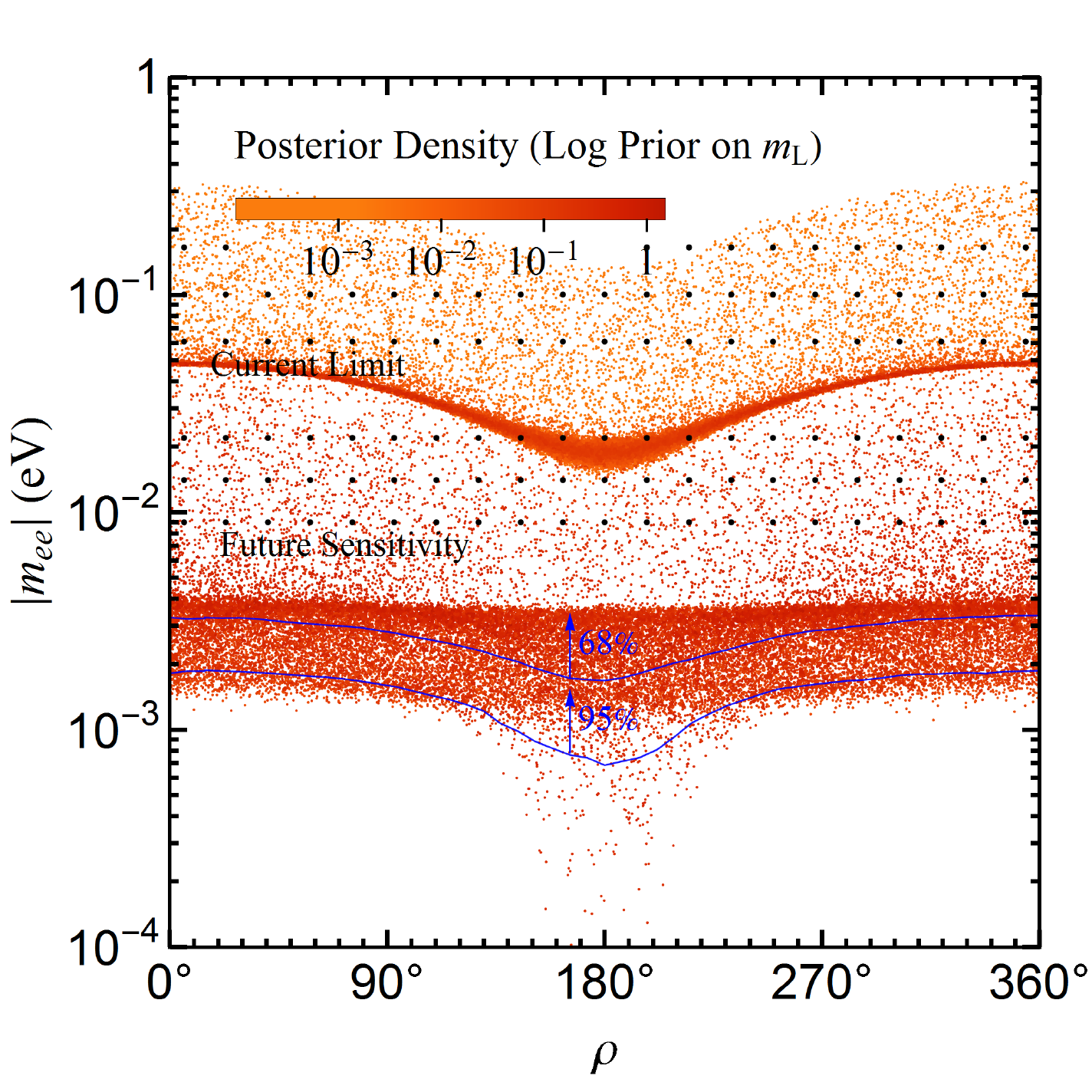} }
\subfigure{
\includegraphics[width=0.48\textwidth]{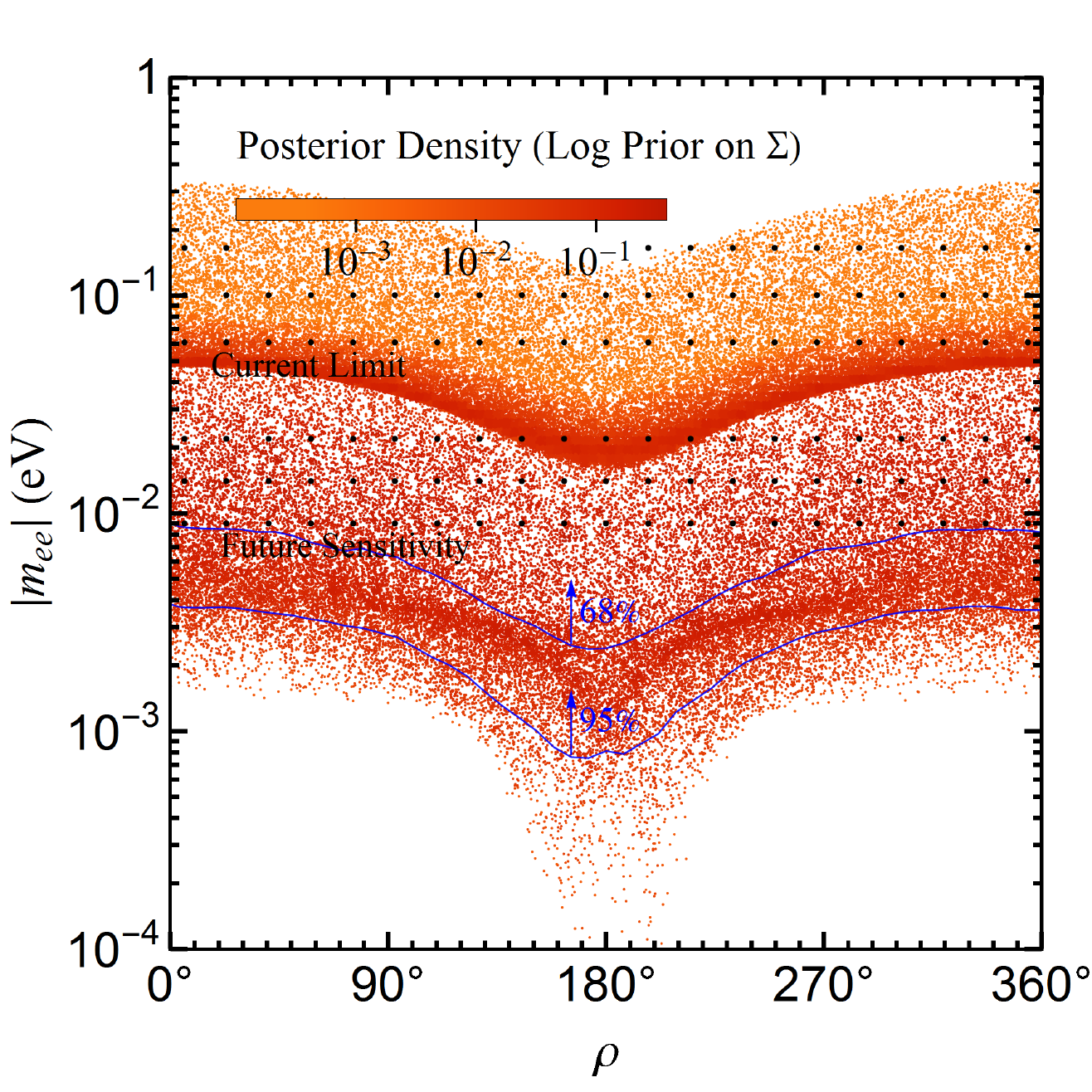} }
\subfigure{
\hspace{-0.2cm}
\includegraphics[width=0.48\textwidth]{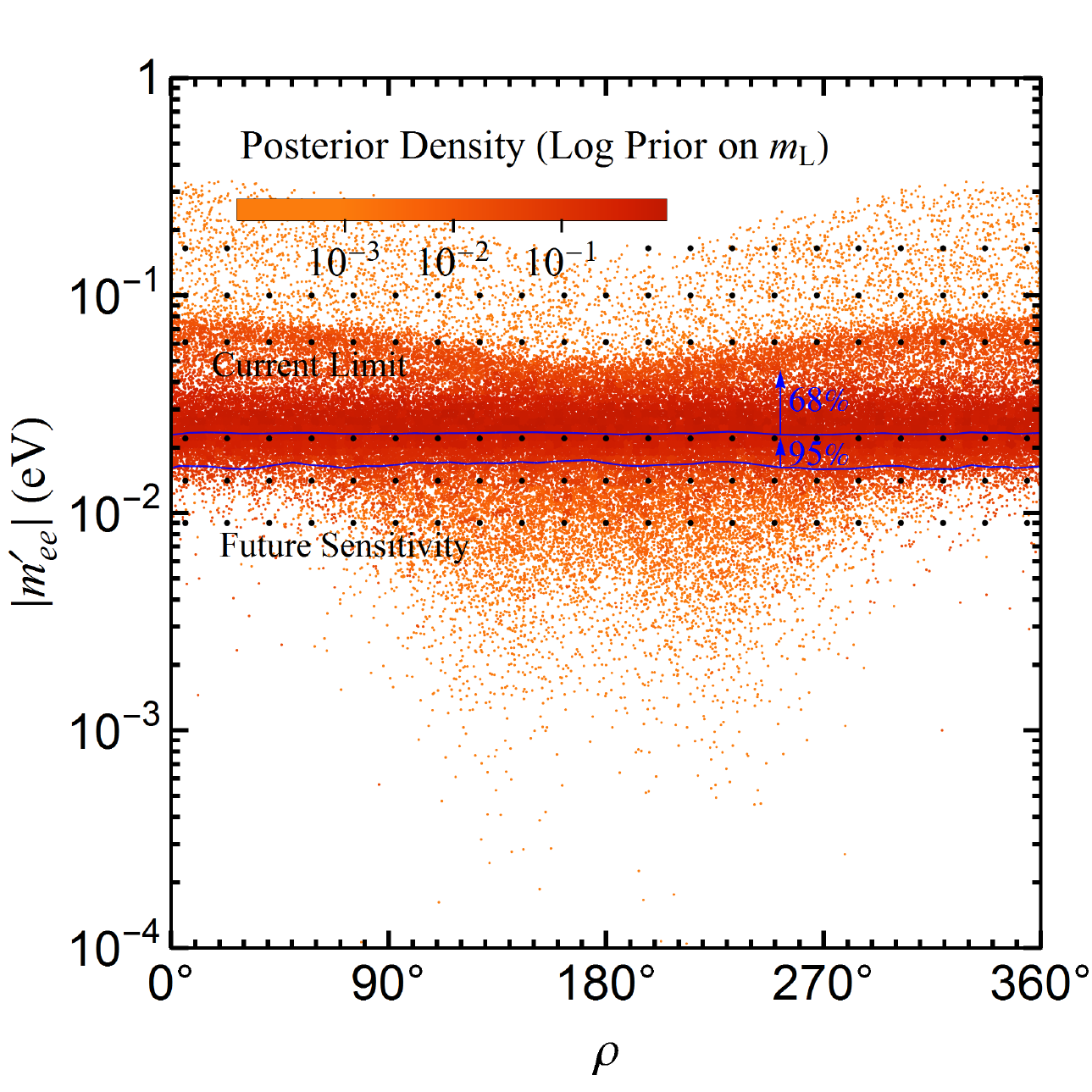} }
\subfigure{
\includegraphics[width=0.48\textwidth]{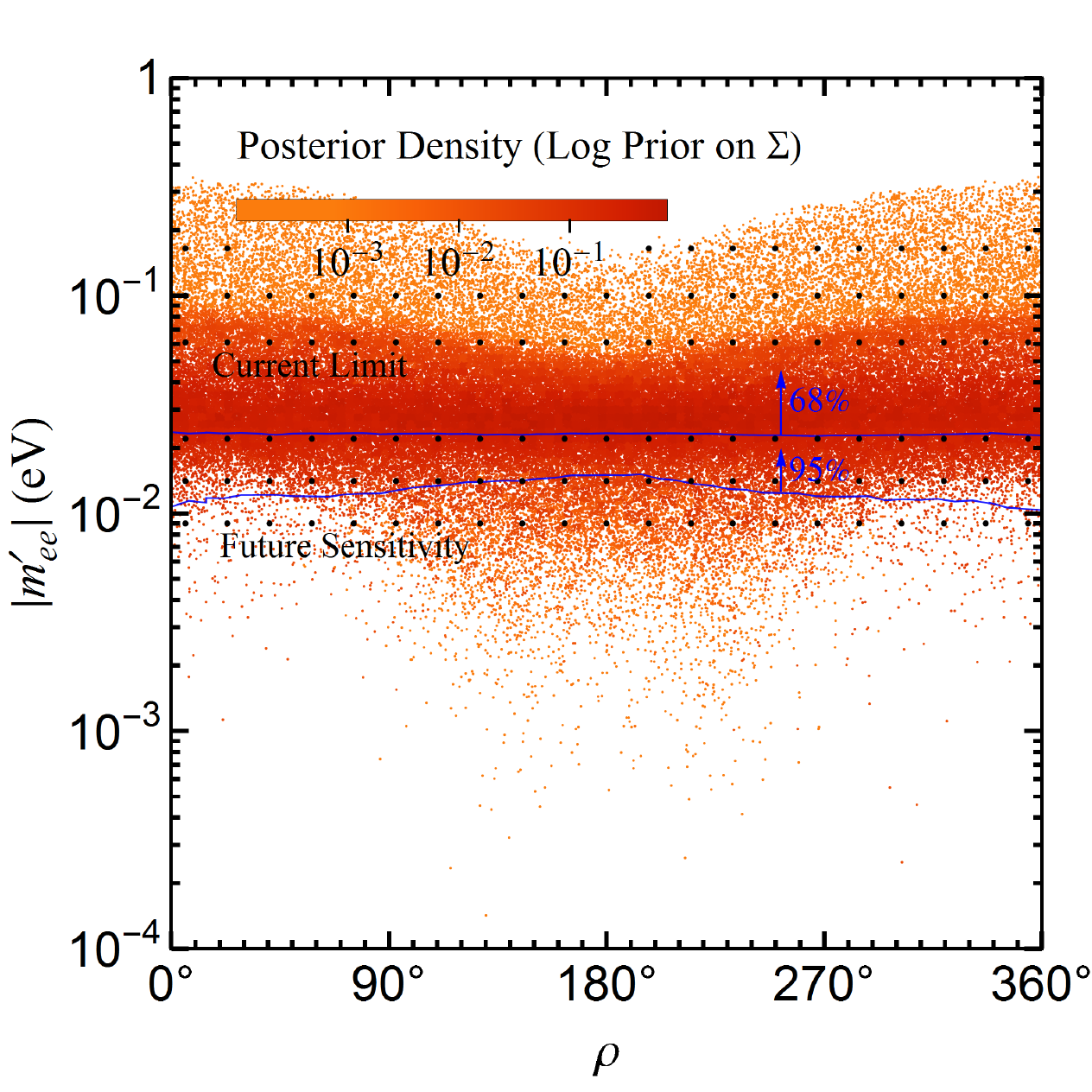} }
\end{center}
\vspace{-0.5cm}
\caption{The posterior probability densities in the $|m^{}_{ee}|$-$\rho$ (the upper row) or $|m^{\prime}_{ee}|$-$\rho$ (the lower row) plane for different choices of models and priors: (i) the standard $3\nu$ mixing scenario with the logarithmic prior on $m^{}_{\rm L}$ (\emph{Upper-left Panel}), (ii) the standard $3\nu$ mixing scenario with the logarithmic prior on $\Sigma$ (\emph{Upper-right Panel}), (iii) the (3+1)$\nu$ mixing scenario with the flat prior on $m^{}_{\rm L}$ (\emph{Lower-left Panel}) and (iv) the (3+1)$\nu$ mixing scenario with the flat prior on $\Sigma$ (\emph{Lower-right Panel}). The notations and conventions are the same as those in Fig.~\ref{fig:3}, except that the lightest neutrino mass $m^{}_{\rm L}$ instead of the Majorana CP phase $\rho$ is marginalized over.}
\label{fig:4}
\end{figure}

After specifying the likelihood functions for the relevant experimental data and fixing the prior probability distributions of model parameters in the previous section, we are ready to compute the posterior distributions of the derived parameters $|m^{}_{ee}|$ and $|m^{\prime}_{ee}|$ by using Eq.~(\ref{eq:Bayesian}). In fact, the posterior probability distribution in Eq.~(\ref{eq:Bayesian}) for the model parameters is calculated via the Monte Carlo sampling, which has been done with the help of the MultiNest routine~\cite{Feroz:2007kg, Feroz:2008xx, Feroz:2013hea}. In Fig.~\ref{fig:3}, we present the posterior sampling distributions in the $|m^{}_{ee}|$-$m^{}_{\rm L}$ plane for the standard 3$\nu$ mixing scenario (the upper row) or in the $|m^{\prime}_{ee}|$-$m^{}_{\rm L}$ plane for the (3+1)$\nu$ mixing scenario (the lower row). The scattered points stand for the sampling data, and one can read off the corresponding posterior probabilities from their colors. Now we explain how to practically do so. For a given point, one can first look at the color legend and find the value of its posterior density, which is denoted as $p$. Then, the posterior probability $P$ can be calculated by definition as the product of $p$ and the area $\mathcal{A}$ of a small region, in which the point is located. For instance, take a small square in the $|m^{}_{ee}|$-$m^{}_{\rm L}$ plane, and its area is thus given by $\mathrm{d}\mathcal{A} \equiv \mathrm{d} \left[\log^{}_{10}(|m^{}_{ee}|/{\rm eV})\right] \times \mathrm{d}\left[\log^{}_{10}(m^{}_{\rm L}/{\rm eV})\right]$. Notice that the total posterior probability is normalized to one for each plot. Several comments on the numerical results in Fig.~\ref{fig:3} are helpful.
\begin{enumerate}
\item In the upper-left panel, the posterior distribution in the $|m^{}_{ee}|$-$m^{}_{\rm L}$ plane is shown for the standard 3$\nu$ mixing scenario, where the logarithmic prior on $m^{}_{\rm L}$ is assumed. The results for the logarithmic prior on $\Sigma$ are plotted in the upper-right panel. In both panels, the thin dot-dashed (or dashed) curves indicate the boundaries of the effective neutrino mass $|m^{}_{ee}|$ in the IO (or NO) case, where the best-fit values of neutrino mixing angles and mass-squared differences are input. Moreover, the current limit (taken from Ref.~\cite{KamLAND-Zen:2016pfg} for the tightest one) on or the future sensitivity (of a ton-scale $0\nu\beta\beta$ decay experiment like nEXO \cite{Gerbino:2016ehw}) to $|m^{}_{ee}|$ is represented by three horizontal dotted lines. The wide range between the upper and lower lines can be ascribed to the NME uncertainty. Comparing the distributions in the left and right panels, one can observe that a larger weight has been given to smaller values of $m^{}_{\rm L}$ in the assumption of a logarithmic prior on $m^{}_{\rm L}$, as already emphasized in the previous section.

\item An urgent question is how likely $|m^{}_{ee}|$ is vanishingly small in the NO case, which has been quantitatively addressed in Refs.~\cite{Agostini:2017jim, Caldwell:2017mqu}. In order to draw a prior-independent conclusion from the posterior distributions, we treat the scenarios with different values of $m^{}_{\rm L}$ as different models. For each fixed $m^{}_{\rm L}$, the posterior distribution of $|m^{}_{ee}|$ can be derived with the help of the likelihood $\mathcal{L}^{}_{ 3\nu}$. Then, one can calculate the probability for the true value of the effective neutrino mass to be above a certain $|m^{}_{ee}|$. The probability contours are plotted as the blue curves in Fig.~\ref{fig:3}, where several representative values, i.e., $68\%$, $95\%$, $99\%$ and $99.7\%$, are shown.  It is evident that the probability for $|m^{}_{ee}|$ to be vanishingly small, e.g., $|m^{}_{ee}| < 10^{-4}~{\rm eV}$, is tiny (less than $0.3\%$). This conclusion is independent of the priors on $m^{}_{\rm L}$, as it should be. In particular, the probability for $|m^{}_{ee}| > 10^{-3}~{\rm eV}$ is larger than $95\%$ even when $m^{}_{\rm L}$ is located in the regime where the destructive cancellation caused by the unknown Majorana CP phases occurs.  

\item In the two panels in the lower row of Fig.~\ref{fig:3}, the posterior probability distributions in the (3+1)$\nu$ mixing scenario have been presented, where the notations and conventions for the curves are the same as those in the plots in the upper row. It is straightforward to observe that the presence of the eV-mass sterile neutrino shifts the effective neutrino mass to higher values. As the future ton-scale $0\nu\beta\beta$ decay experiments are able to explore the effective neutrino mass to the level of $\mathcal{O}(10^{-2})~{\rm eV}$, the inclusion of the sterile neutrino can raise the effective mass to the level that is within the reach of the next-generation experiments even for a very small $m^{}_{\rm L}$. If the sensitivity at the $\mathcal{O}(10^{-2})~{\rm eV}$ level is achieved, more than $99.7\%$ of the region of $|m^{\prime}_{ee}|$ can be covered for $m^{}_{\rm L} \lesssim 10^{-2}~{\rm eV}$. When $m^{}_{\rm L} \gtrsim 10^{-2}~{\rm eV}$, the chance for $|m^{\prime}_{ee}|$ to fall into the cancellation region increases. However, even in this case, at least $95\%$ of the $|m^\prime_{ee}|$ range can be probed. Therefore, in the statistical sense, it is quite promising to check the (3+1)$\nu$ mixing scenario with an eV-mass sterile neutrino in the future $0\nu\beta\beta$ decay experiments. 
\end{enumerate}

In Fig.~\ref{fig:4}, we present the posterior distributions in the $|m^{}_{ee}|$-$\rho$ (the upper row) or $|m^\prime_{ee}|$-$\rho$ plane (the lower row) by marginalizing over the lightest neutrino mass $m^{}_{\rm L}$ instead of the Majorana CP phase $\rho$. The notations and conventions are the same as those in Fig.~\ref{fig:3}. The area in the $|m^{}_{ee}|$-$\rho$ plane is defined as $\mathrm{d}\mathcal{A} \equiv \mathrm{d} \left[\log^{}_{10}(|m^{}_{ee}|/{\rm eV})\right] \times \mathrm{d}\left[\rho/{\rm rad}\right]$ in the $3\nu$ mixing scenario, and likewise for the (3+1)$\nu$ mixing scenario. Now the blue solid curves in Fig.~\ref{fig:4} stand for the contours of the probability for the effective neutrino mass to be above a certain $|m^{}_{ee}|$ or $|m^\prime_{ee}|$. These contours become dependent on the $m^{}_{\rm L}$ priors, because the prior information of $m^{}_{\rm L}$ has been integrated into the posterior distribution. It is worthwhile to notice that the dependence of posterior distributions on $\rho$ is very weak for the (3+1)$\nu$ mixing scenario. In the $3\nu$ mixing scenario, the fine structure around $\rho \approx \pi$ due to the cancellation can be observed. Therefore, it seems difficult to determine the Majorana CP phase $\rho$ if $|m^{}_{ee}|$ takes the value far away from that in the cancellation region. 

\begin{figure}[t!]
	\begin{center}
		\subfigure{
			\hspace{-0.2cm}
			\includegraphics[width=0.48\textwidth]{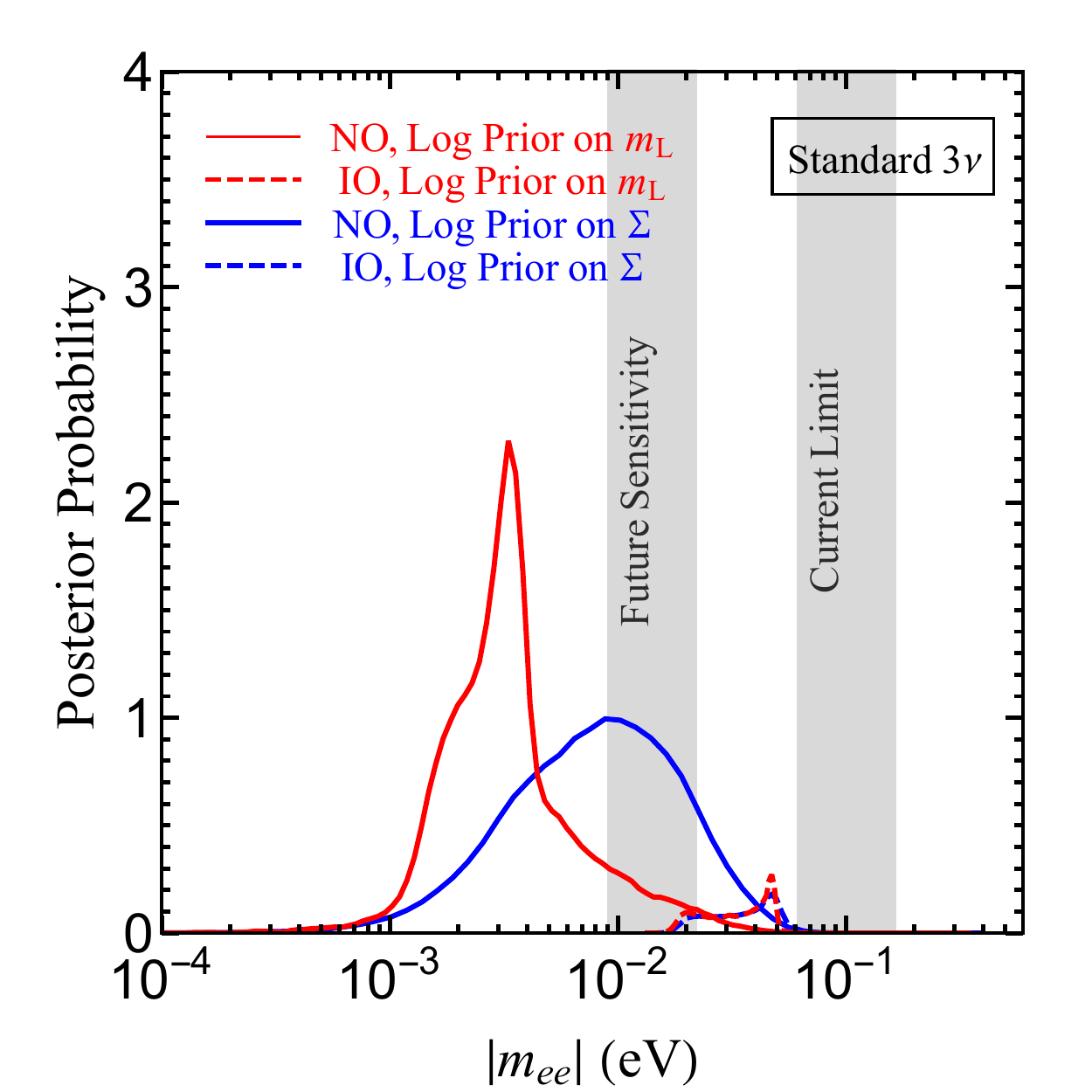} }
		\subfigure{
			\includegraphics[width=0.48\textwidth]{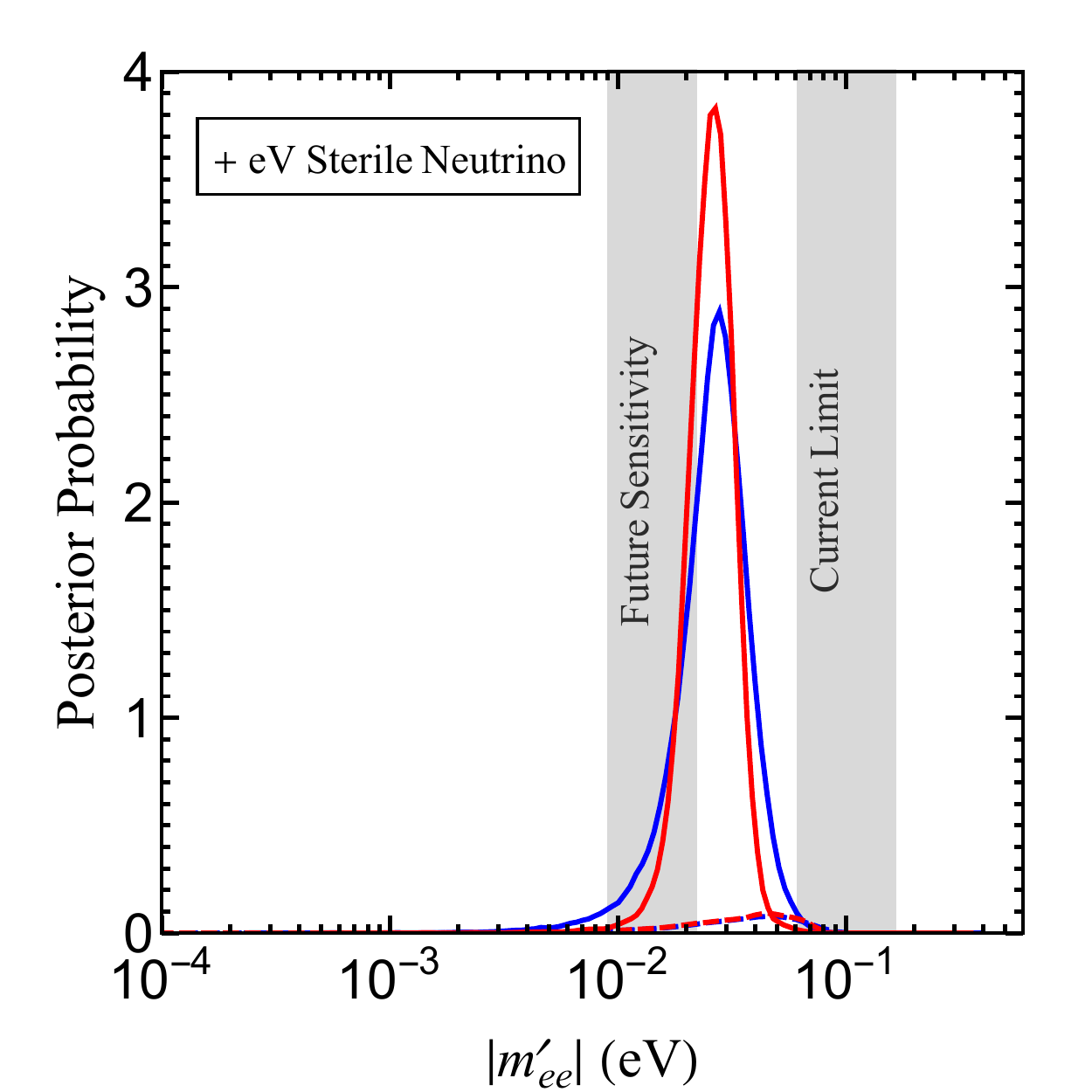} }
	\end{center}
	\vspace{-0.5cm}
	\caption{The posterior probability densities of $|m^{}_{ee}|$ or $|m^{\prime}_{ee}|$ for different choices of models and priors. The standard $3\nu$ results are presented in the left panel, while those with the sterile neutrino in the right panel. The posteriors with the logarithmic prior on $m^{}_{\rm L}$ (the logarithmic prior on $\Sigma$) are plotted as the red (blue) solid curves in the NO case, but as the dashed curves in the IO case.}
	\label{fig:5}
\end{figure}

As the effective neutrino mass can be directly extracted from the experimental data on $0\nu\beta\beta$ decays, it is interesting to see the posterior distribution of $|m^{}_{ee}|$ or $|m^\prime_{ee}|$, which can be obtained by marginalizing over both $m^{}_{\rm L}$ and $\rho$. The final results can be found in Fig.~\ref{fig:5}. For the standard $3\nu$ case in the left panel, if we choose the logarithmic prior on $m^{}_{\rm L}$ for NO (red solid curve), a large fraction (about $92\%$) of the probable range of $|m^{}_{ee}|$ is unreachable for the future ton-scale $0\nu \beta\beta$ decay experiments. With a logarithmic prior on $\Sigma$ (blue solid curve), the next generation experiments can cover about $ 41\%$ of the range. As we have observed before, adding an eV-mass sterile neutrino can greatly enhance the probability of the effective neutrino mass $|m^\prime_{ee}|$ to larger values. The future $0\nu\beta\beta$ decay experiments with a sensitivity to the effective neutrino mass of $\mathcal{O}(10^{-2})~{\rm eV}$ can cover around $99.4\%$ ($97.4\%$) of the posterior space for the logarithmic prior on $m^{}_{\rm L}$ (the logarithmic prior on $\Sigma$) in the (3+1)$\nu$ mixing scenario. According to the posterior distributions in Fig.~\ref{fig:5}, we find that the average value of the effective neutrino mass is shifted from $\overline{|m^{}_{ee}|} = 3.37\times 10^{-3}~{\rm eV}$ (or $7.71\times 10^{-3}~{\rm eV}$) in the standard $3\nu$ mixing scenario to $\overline{|m^{\prime}_{ee}|}=2.54\times 10^{-2}~{\rm eV}$ (or $2.56\times 10^{-2}~{\rm eV}$) in the (3+1)$\nu$ mixing scenario, with the logarithmic prior on $m^{}_{\rm L}$ (or on $\Sigma$). Therefore, a null signal from the future $0\nu\beta\beta$ decay experiments will be able to set a very stringent constraint on the sterile neutrino mass and mixing angle.

\section{Concluding Remarks}

In this short note, we have carried out a Bayesian analysis of the effective neutrino mass in the $0\nu\beta\beta$ decays in both the standard $3\nu$ mixing scenario and the (3+1)$\nu$ mixing scenario. With the latest experimental information, including the global-fit analysis of neutrino oscillation data, the cosmological observations from the {\it Planck} satellite and the current limits from the $0\nu\beta\beta$ decay experiments, the posterior probability distributions of the effective neutrino mass $|m^{}_{ee}|$ in the standard $3\nu$ mixing scenario and $|m^\prime_{ee}|$ in the (3+1)$\nu$ mixing scenario have been updated. 

Our main results of the posterior distributions have been summarized in Fig.~\ref{fig:3} and Fig.~\ref{fig:5}. Adding an eV-mass sterile neutrino slightly mixing with ordinary neutrinos is likely to enhance the effective neutrino mass to the level of ${\cal O}(10^{-2})~{\rm eV}$, which is within the reach of the next generation $0\nu\beta\beta$ decay experiments, regardless of the prior information on the absolute mass scale of ordinary neutrinos. In other words, if a null signal is observed in future ton-scale  $0\nu\beta\beta$ decay experiments, we can place very strong limits on the parameter space of the (3+1)$\nu$ mixing scenario, assuming that massive neutrinos are of Majorana nature. The sensitivity of future $0\nu\beta\beta$ decay experiments to the sterile neutrino mass and mixing angle deserves a dedicated study, which will be left for the upcoming works.

\section*{Acknowledgements}

This work was supported in part by the National Natural Science Foundation of China under grant No.~11775232 and No.~11835013, and by the CAS Center for Excellence in Particle Physics.

\end{document}